\documentclass{PoS}
\usepackage[sort&compress]{natbib}
\bibpunct{[}{]}{,}{n}{}{}

\newcommand{\comment}[1]{}
\newcommand{\lr}[1]{ \left( #1 \right) }
\newcommand{\lrs}[1]{ \left[ #1 \right] }

\newcommand{\tr}{ {\rm Tr} \: }

\newcommand{\expa}[1]{ \exp{\left( #1 \right)} }

\title{Rigidity and percolation of center vortices}

\ShortTitle{Rigidity and percolation of center vortices}

\author{\speaker{M. I. Polikarpov}, V. I. Zakharov\\
        ITEP, Moscow, Russia\\
        E-mail: \email{polykarp@itep.ru}, \email{xxz@mppmu.mpg.de}}

\author{P. V. Buividovich\\
 JIPNR "Sosny", Minsk, Belarus\\
 E-mail: \email{buividovich@tut.by}}

\abstract{Effective action of center vortices in $SU\lr{2}$ lattice gauge theory is investigated by studying the correlation between the action density on their worldsheets and their geometric properties. It turns out that center vortices are rigid, however, their dynamics is more complicated than that of rigid random surfaces, since some coupling constants have nonstandard scaling dimensions. As a result, the properties of center vortices are almost completely determined by curvature-dependent effects. This, in turn, provides a qualitative explanation of vortex percolation.}

\FullConference{The XXV International Symposium on Lattice Field Theory\\
                 July 30 - August 4 2007\\
                 Regensburg, Germany}

\begin{document}
\sloppy

\section{Introduction}
\label{sec:Introduction}

 Since the seminal work of t'Hooft \cite{tHooft:78} center vortices in non-Abelian gauge theories are often discussed as field configurations which are responsible for confinement. During the last decade the properties of center vortices were extensively studied using lattice simulations, and their importance for the low-energy behavior of Yang-Mills theories has been fully verified: it was found that confinement and chiral symmetry breaking are absent if vacuum expectation values are calculated for lattice configurations without center vortices \cite{DelDebbio:98}. In \cite{Polikarpov:03:1, DelDebbio:98} it was demonstrated that the area of center vortices in $SU\lr{2}$ lattice gauge theory scales in physical units of length, which gives one more argument that center vortices are not just lattice artifacts. Many other interesting properties of center vortices were discovered in \cite{Polikarpov:03:1, Polikarpov:03:2}. For instance, it was found that center vortices are genuinely thin and carry quadratically divergent action density. It also turned out that vortex worldsheets are densely covered with Abelian monopoles \cite{Polikarpov:03:2}. However, a proper theoretical description of vortex dynamics is still to be found.

 In recent works \cite{Polikarpov:03:1, Buividovich:07:3} an attempt has been made to find an effective vortex action by studying the correlation between the action density on vortex worldsheets and their local geometry. Center vortices turned out to be rigid, i.e. extra energy is required to bend the vortex. It was conjectured that rigidity may be induced by some two-dimensional fields localized on vortices. Rigidity of center vortices allows one to understand at least qualitatively the observed scaling of vortex area, since in contrast to the simplest case of random surfaces with Nambu-Goto action rigid random surfaces may have a nontrivial continuum limit \cite{Ambjorn:94:1}. In fact rigidity is a common feature of any vortex-like solution in gauge theories, since the terms in the string action which make vortex worldsheets rigid are the first terms in the general gradient expansion of the effective vortex action \cite{Engelhardt:00:2}.

 On the other hand, there are some important properties of center vortices which are not so easily explained by the model of rigid strings. One of them is the percolation of vortex worldsheets, which can play a crucial role in the confinement mechanism (for instance, it is known that percolating vortex disappears at the deconfinement phase transition \cite{Chernodub:07:1}). The aim of this paper is to investigate the effective vortex action in more details than in \cite{Buividovich:07:3} and to discuss how such action can describe the emergence of percolating vortex clusters.

 %It turns out that the divergent part of the effective vortex action is almost completely saturated by curvature-dependent terms, which means that vortices can be effectively described as tensionless rigid strings. This action also appears to be unstable with respect to the growth of the genus of vortex worldsheets, which can qualitatively explain percolation of vortices. Such instability indicates that the effective vortex action may be nonrenormalizable.

\section{Effective vortex action}
\label{sec:eff_vort_act}

 A basic assumption of \cite{Buividovich:07:3, Polikarpov:03:1} and this work is that the average total excess of action on the vortex worldsheets is equal to the effective vortex action $W_{eff}\lrs{\Sigma}$ obtained by integrating over all degrees of freedom in the theory except the geometry of the vortex worldsheet $\Sigma$. In general, $W_{eff}\lrs{\Sigma}$ can include nonlocal terms, for instance, of the form $R \Delta^{-1} R$, where $R$ is the internal curvature of the vortex worldsheet and $\Delta$ is the covariant two-dimensional Laplacian. As it is very difficult to trace such nonlocal terms on presently available lattices, in this paper we will assume that $W_{eff}\lrs{\Sigma}$ is indeed local. Furthermore, we will simplify our task by assuming that $W_{eff}\lrs{\Sigma}$ depends only on the simplest local two-dimensional geometric invariants -- internal curvature $R$ and extrinsic curvature $K$, i.e. $W_{eff}\lrs{\Sigma} = \int \limits_{\Sigma} d^{2}\xi \sqrt{g} \: w_{eff}\lr{K,R; a}$, where $\xi^{b}$, $b = 1,2$ are the coordinates on the vortex worldsheet, $g_{ab} = \partial_{a} x^{\mu} \partial_{b} x^{\mu}$ is the induced metric and $g$ is its determinant. The extrinsic curvature $K$ can be written in the following form:
\begin{eqnarray}
\label{ExtCurvDef}
K = g^{a b} \partial_{a} t^{\mu \nu} \partial_{b} t^{\mu \nu} = - t^{\mu \nu} \Delta t^{\mu \nu}
\end{eqnarray}
where $t^{\mu \nu} = \epsilon^{a b} \: \partial_{a} x^{\mu} \partial_{b} x^{\nu}$ is the two-form tangent to the vortex worldsheet $\Sigma$. We has also taken into account that the effective action should depend on the lattice spacing $a$, according to the Wilson's renormalization group equations.

\begin{figure}[ht]
  \includegraphics[width=5cm, angle=-90]{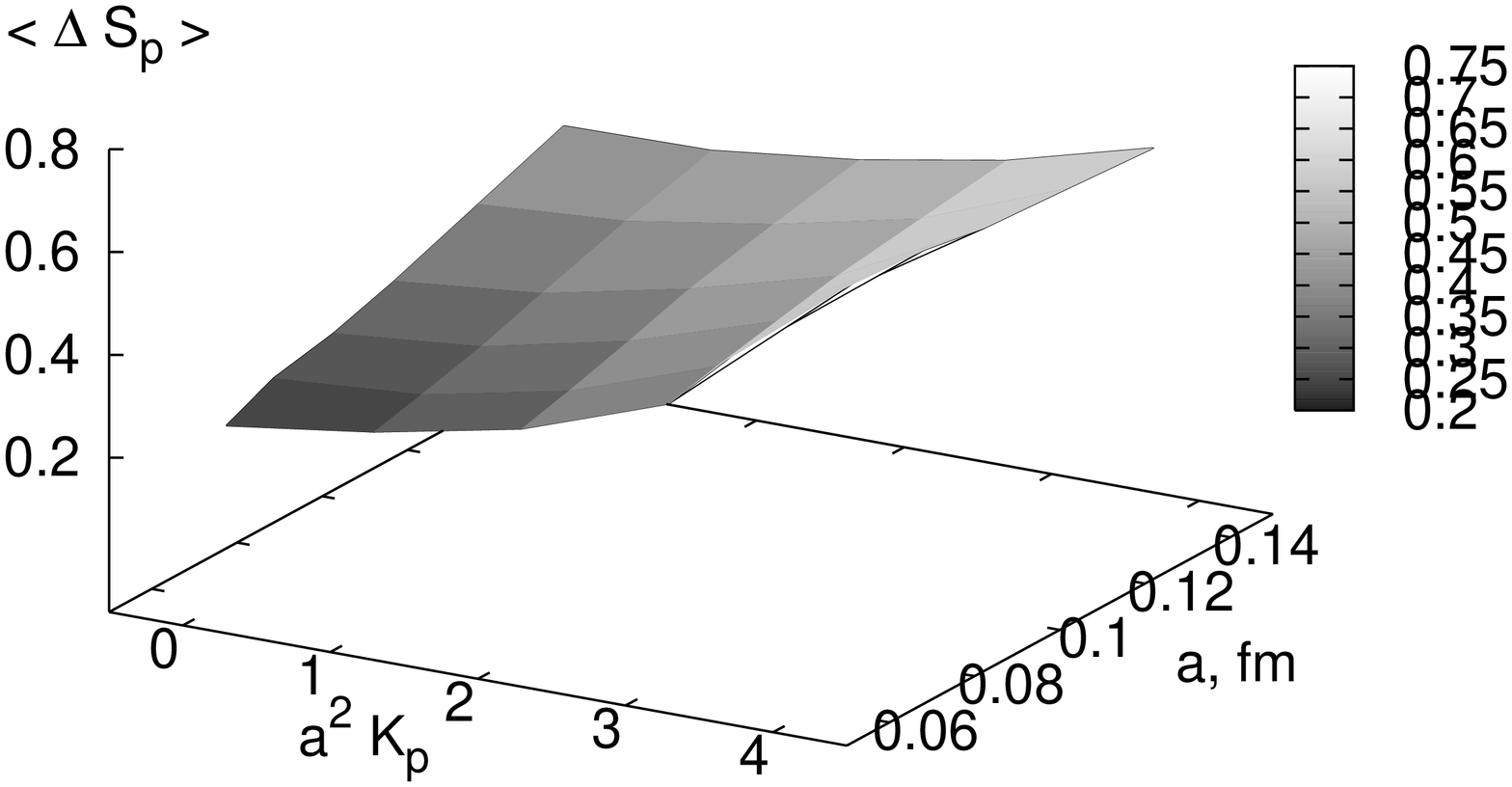}
  \includegraphics[width=5cm, angle=-90]{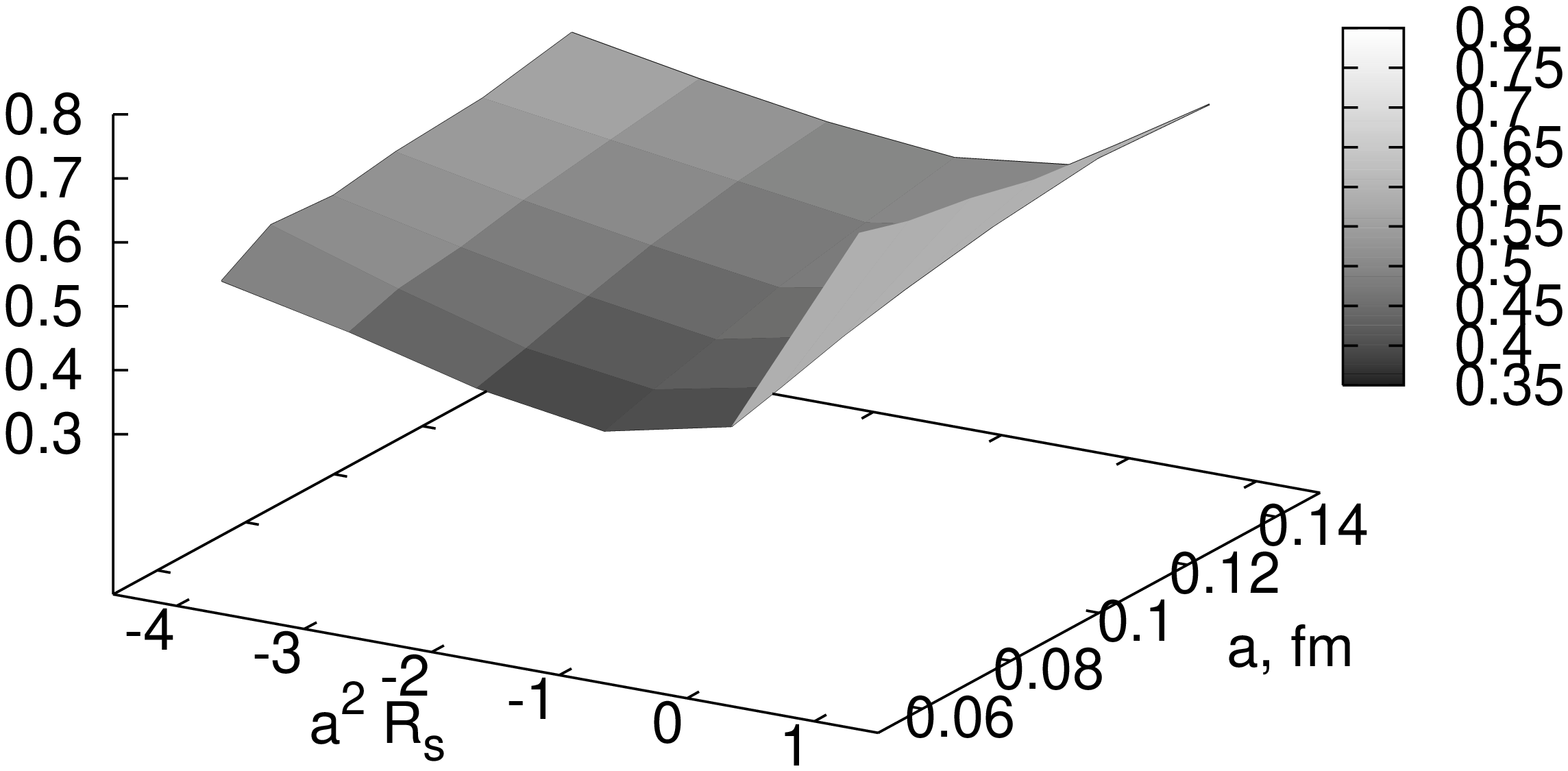}\\
  \caption{Average excess of action as the function of extrinsic (left) and internal (right) curvatures and lattice spacing}
  \label{fig:act_vs_curv_spacing}
\end{figure}

 In order to study the correlation between the excess of action density and internal and extrinsic curvatures of vortex worldsheets in pure $SU\lr{2}$ Yang-Mills theory, the excess of action per plaquette $\Delta S_{p} = a^{2} w_{eff}\lr{K,R; a} = \beta \lr{1 - \frac{1}{2} \tr U_{p} } - \langle \beta \lr{1 - \frac{1}{2} \tr U_{p} } \rangle  = \frac{\beta}{2} \lr{ \langle \tr U_{p} \rangle - \tr U_{p}}$ was averaged separately for vortex plaquettes and sites with different values of $K$ and $R$. Lattice discretization of (\ref{ExtCurvDef}) can be written as:
\begin{eqnarray}
\label{ExtCurvDefLat}
a^{2} K_{p} = - t_{p}^{\mu \nu} a^{2} \Delta t_{p}^{\mu \nu}
 =
\sum \limits_{p'} t_{p}^{\mu \nu} \lr{t_{p}^{\mu \nu} - t_{p'}^{\mu \nu}}
 = 4 - n_{p}
\end{eqnarray}
where $\sum \limits_{p'}$ denotes summation over all plaquettes $p'$ adjacent to the plaquette $p$,  $n_{p}$ is the number of vortex plaquettes which are adjacent to $p$ and are parallel to it and the factor $a^{2}$ comes from the definition of derivatives on the lattice. According to (\ref{ExtCurvDefLat}), $a^{2} K_{p}$ takes integer values from $0$ to $4$. Internal curvature for lattice surfaces was defined as in \cite{Buividovich:07:3, Ambjorn:94:1}: $a^{2} R_{s} = 4 - n_{s}$, where $n_{s}$ is the number of sites on the vortex worldsheet which are adjacent to $s$. Thus in contrast to \cite{Buividovich:07:3}, where both $K$ and $R$ were associated with lattice sites, here $K$ is associated with vortex plaquettes and $R$ -- with lattice sites on vortex worldsheets, which yields more self-consistent results.

\begin{figure}[ht]
  \includegraphics[width=5cm, angle=-90]{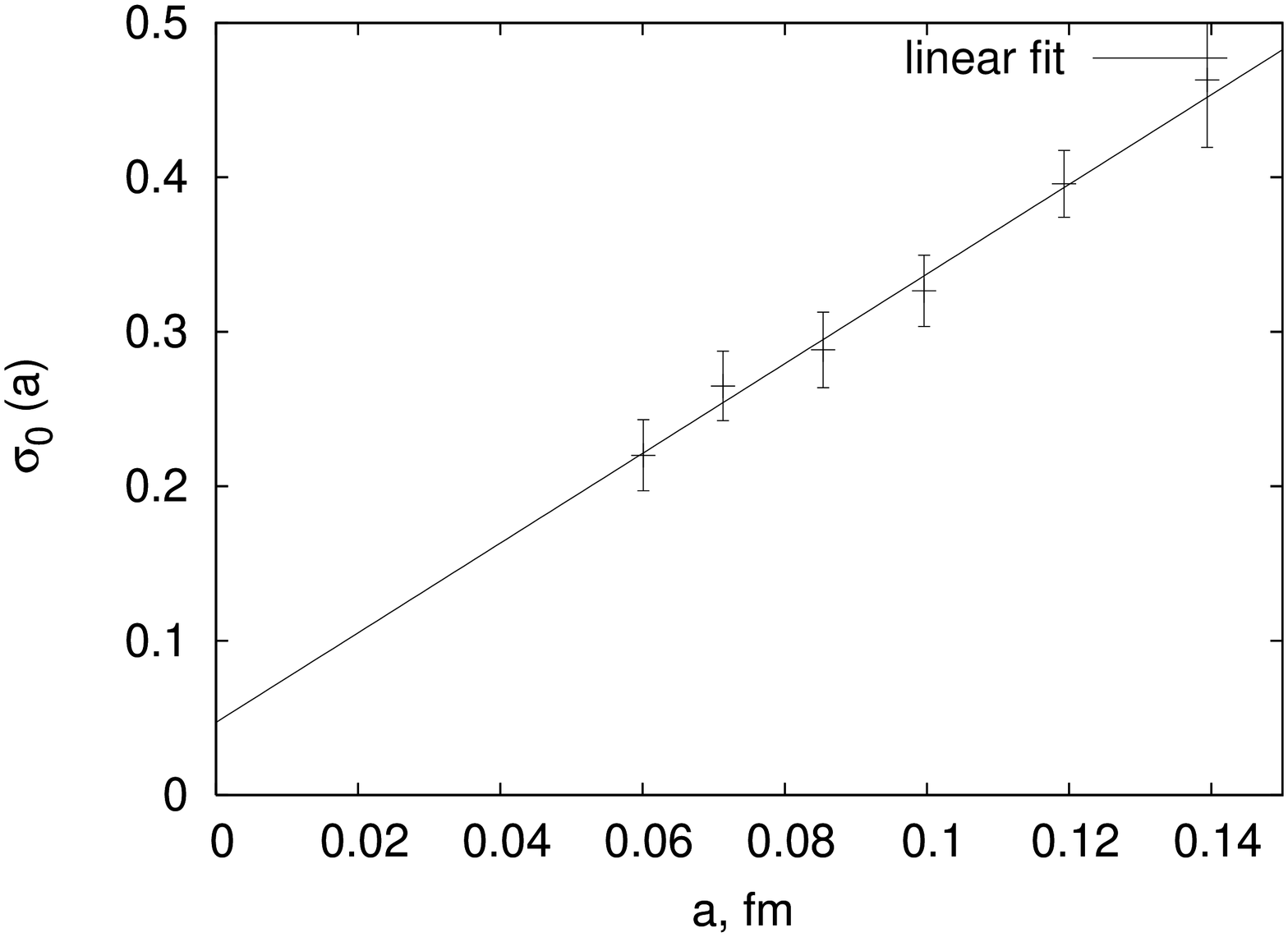}
  \includegraphics[width=5cm, angle=-90]{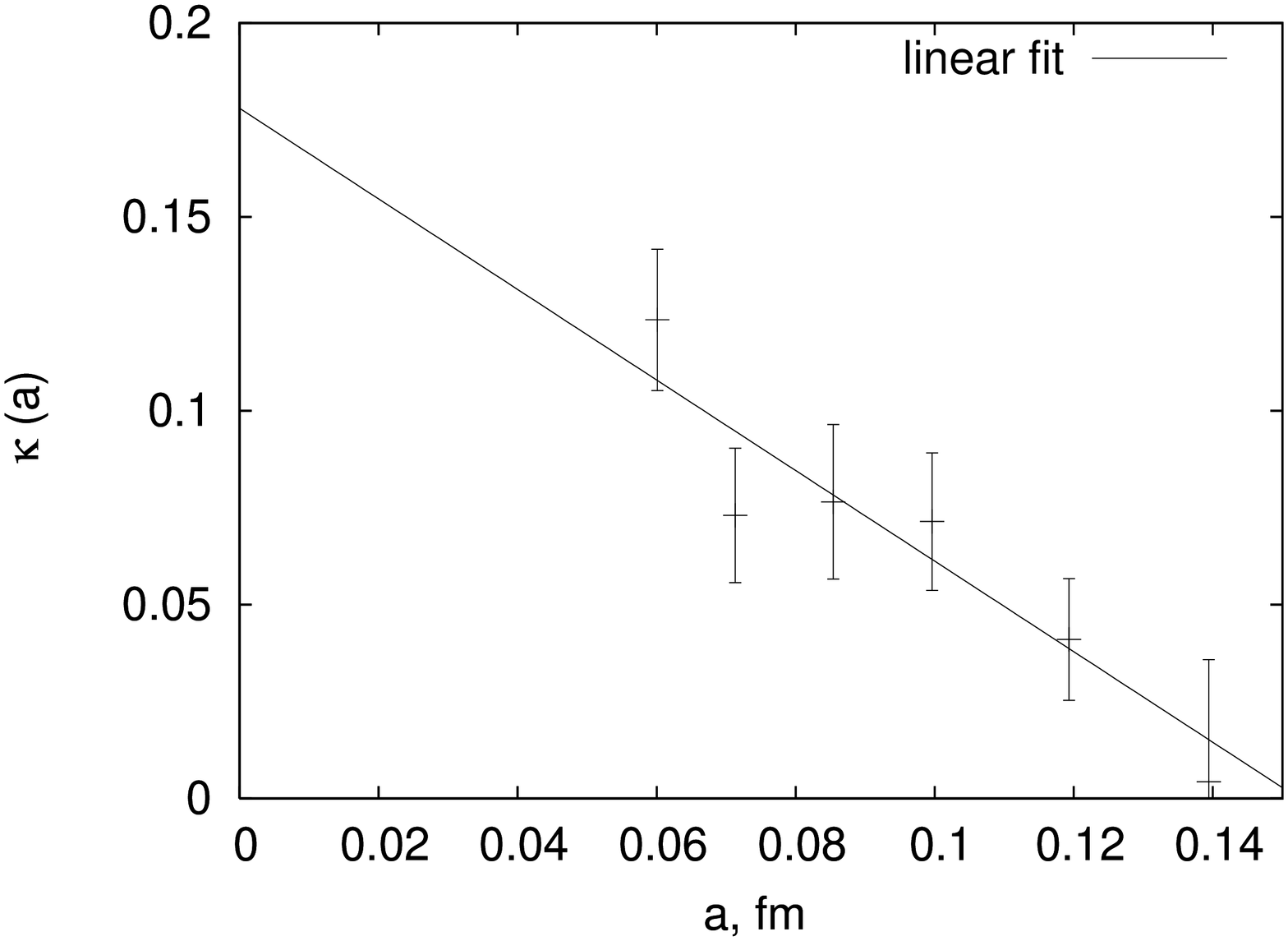}\\
  \includegraphics[width=5cm, angle=-90]{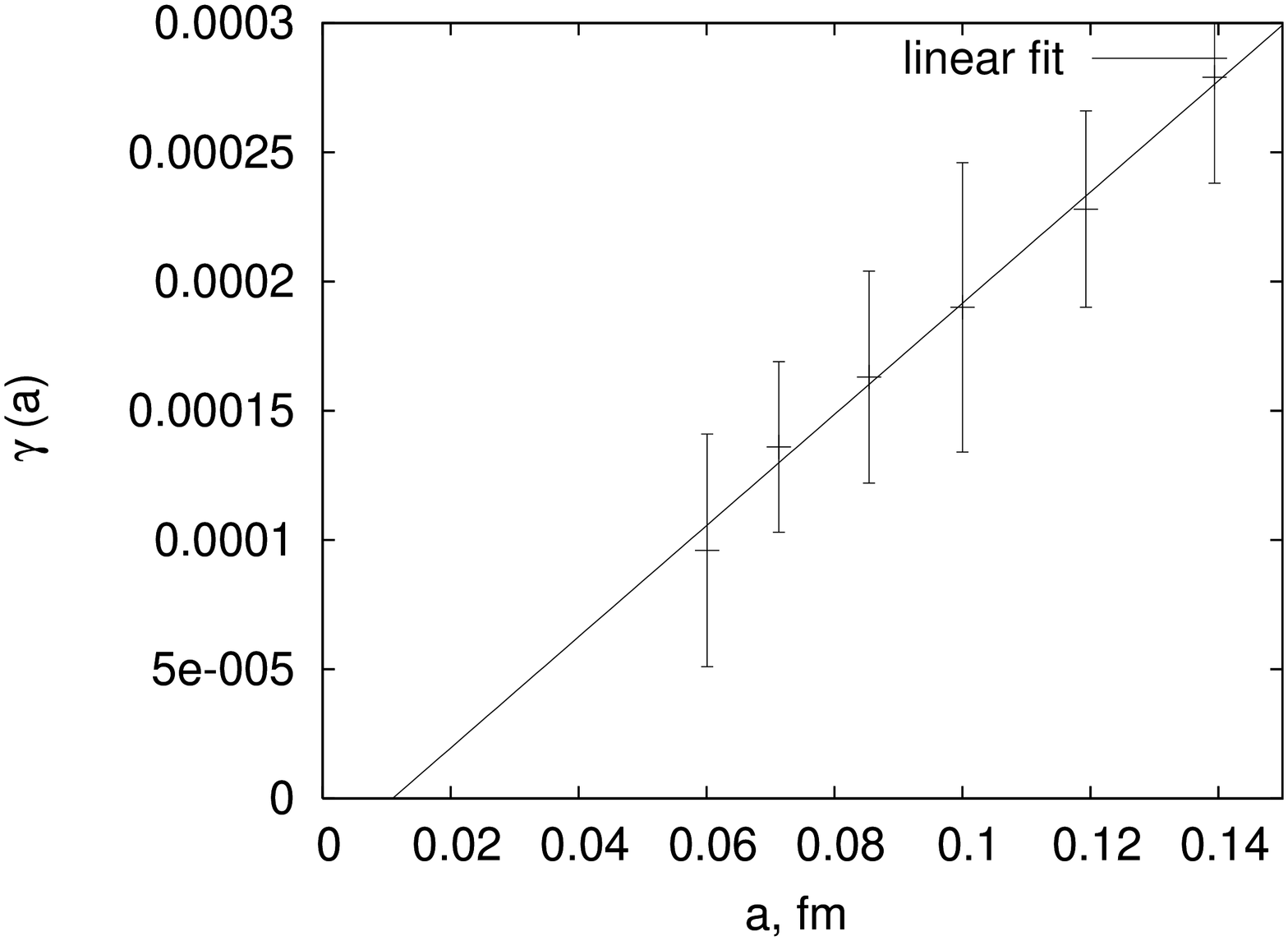}
  \includegraphics[width=5cm, angle=-90]{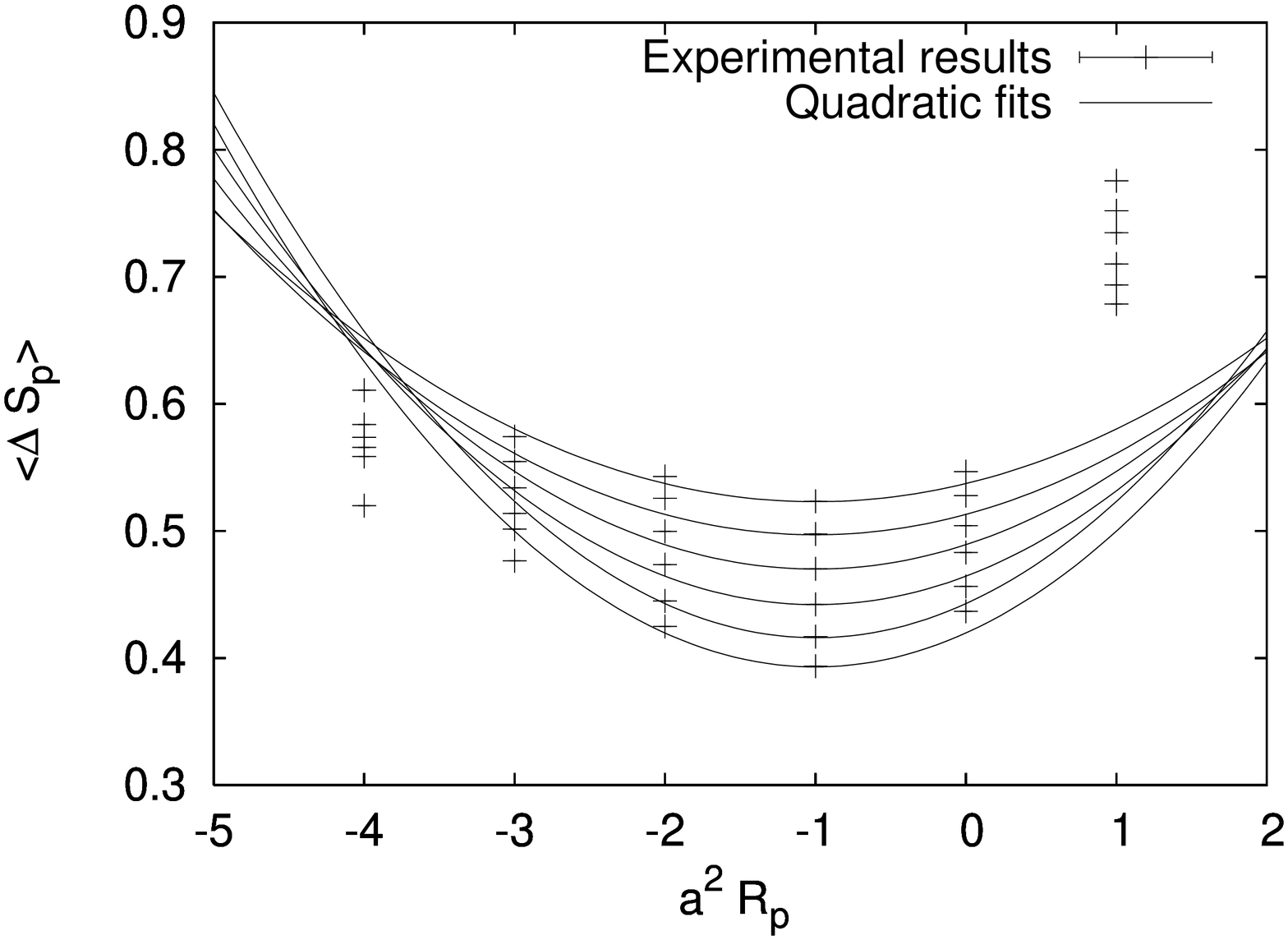}\\
  \caption{
  At upper left: bare vortex tension $\sigma_{0} \lr{a}$, at upper right: coupling constant $\kappa \lr{a}$, at lower left: coupling constant $\gamma\lr{a}$, at lower right: average excess of action per plaquette as the function of  internal curvature in lattice units $a^{2} R_{p}$ at different values of lattice spacing (from $a = 0.14 \: fm$  for the uppermost curve to $a = 0.06 \: fm$ for the lowest curve. The errors in experimental results are not shown in order not to clutter the plot).
  }
  \label{fig:couplings_vs_spacing}
\end{figure}

 Average excess of action per vortex plaquette $\langle \: \Delta S_{p} \: \rangle$ as the function of lattice spacing and extrinsic and internal curvatures (in lattice units) is plotted on Fig. \ref{fig:act_vs_curv_spacing}. Empirically it was found that $w_{eff}\lr{K,R; a}$ is almost linear in $a^{2} K_{p}$ and has a distinct minimum with respect to $a^{2} R_{s}$, which is situated near $a^{2} R_{s} \approx -1$. For the sake of simplicity we have approximated the dependence of $\Delta S_{p}$ on $a^{2} R_{s}$ by a quadratic function, since in fact only the existence of this minimum is essential. The actual dependence of $\Delta S_{p}$ on $a^{2} R_{s}$ at different lattice spacings and the quadratic functions which approximate it are plotted on Fig. \ref{fig:couplings_vs_spacing} at lower right. Thus we assume that $w_{eff}\lr{K,R; a}$ has the following form:
\begin{eqnarray}
\label{EffVortActFit}
w_{eff}\lr{K,R; a} = \sigma\lr{a} + \kappa\lr{a} K + \gamma\lr{a} \: \lr{R - R_{0}\lr{a}}^{2}
\end{eqnarray}
where $R_{0}\lr{a} \approx -1 \cdot a^{-2}$. This action differs from the action of rigid strings used in \cite{Buividovich:07:3} by the terms quadratic in $R$. Although the standard dimensional analysis shows that only $\sigma\lr{a}$ and the terms linear in $K$ and $R$ should survive in the continuum limit, we shall see that for center vortices higher-order terms can be equally important. It is also more convenient to redefine $\sigma\lr{a} = a^{-2} \sigma_{0}\lr{a} = \Lambda_{UV}^{2} \sigma_{0}\lr{a}$. The coefficients $\sigma_{0}\lr{a}$, $\kappa\lr{a}$ and  $\gamma\lr{a}$ as the functions of lattice spacing are plotted on Fig. \ref{fig:couplings_vs_spacing}.

\section{Scaling of coupling constants and renormalization of the effective vortex action}
\label{sec:scaling}

 Dimensional analysis shows that as the continuum limit $a \rightarrow 0$ is approached, the bare vortex tension $\sigma\lr{a}$ should grow as $a^{-2} \sim \Lambda_{UV}^{2}$ and the coupling constant $\gamma\lr{a}$ should tend to zero as $a^{2}$. Since the external curvature $K$ is a marginal operator, the coupling constant $\kappa\lr{a}$ can grow not faster than $\ln a$.

\begin{figure}[ht]
  \includegraphics[width=5cm, angle=-90]{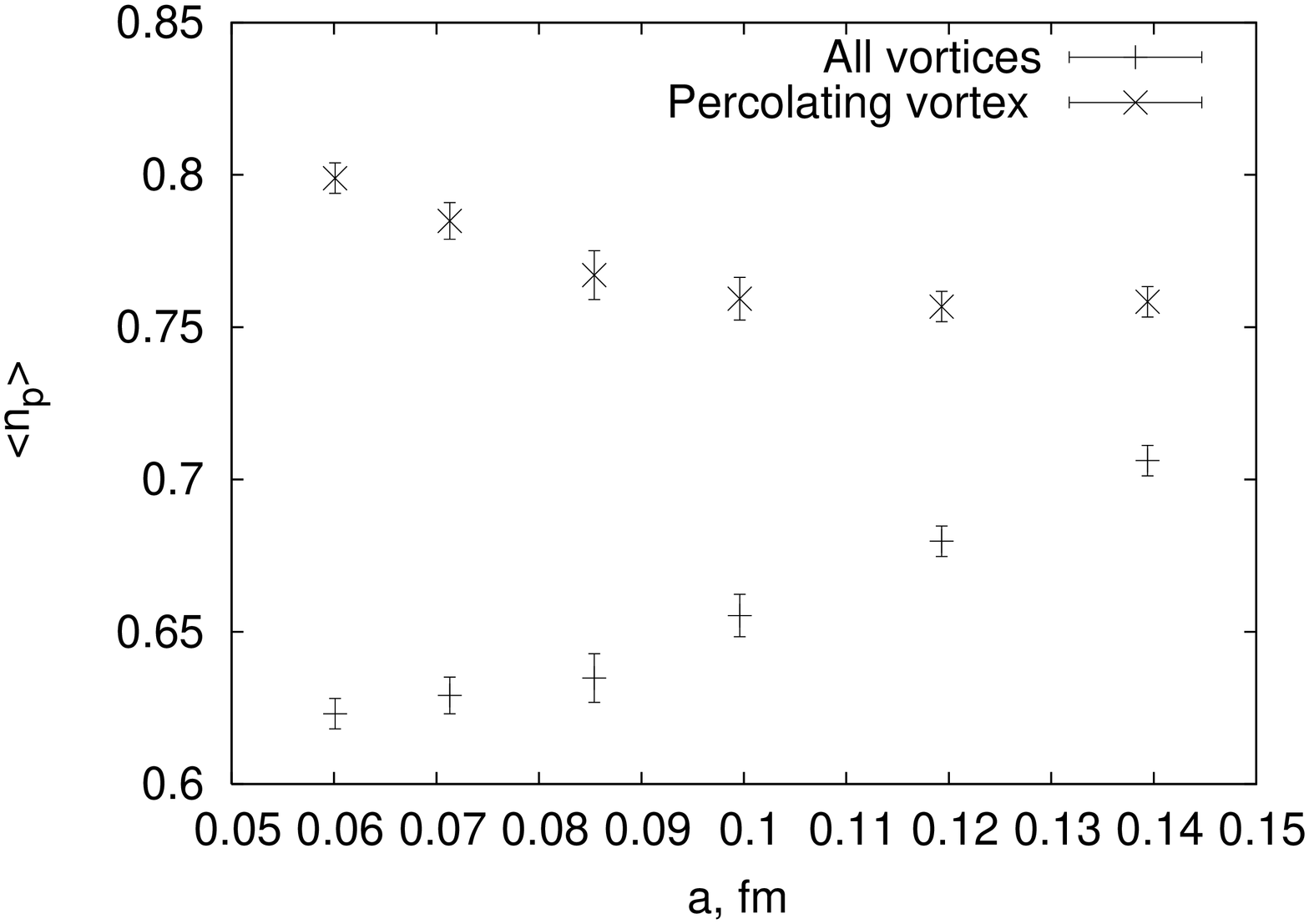}
  \includegraphics[width=5cm, angle=-90]{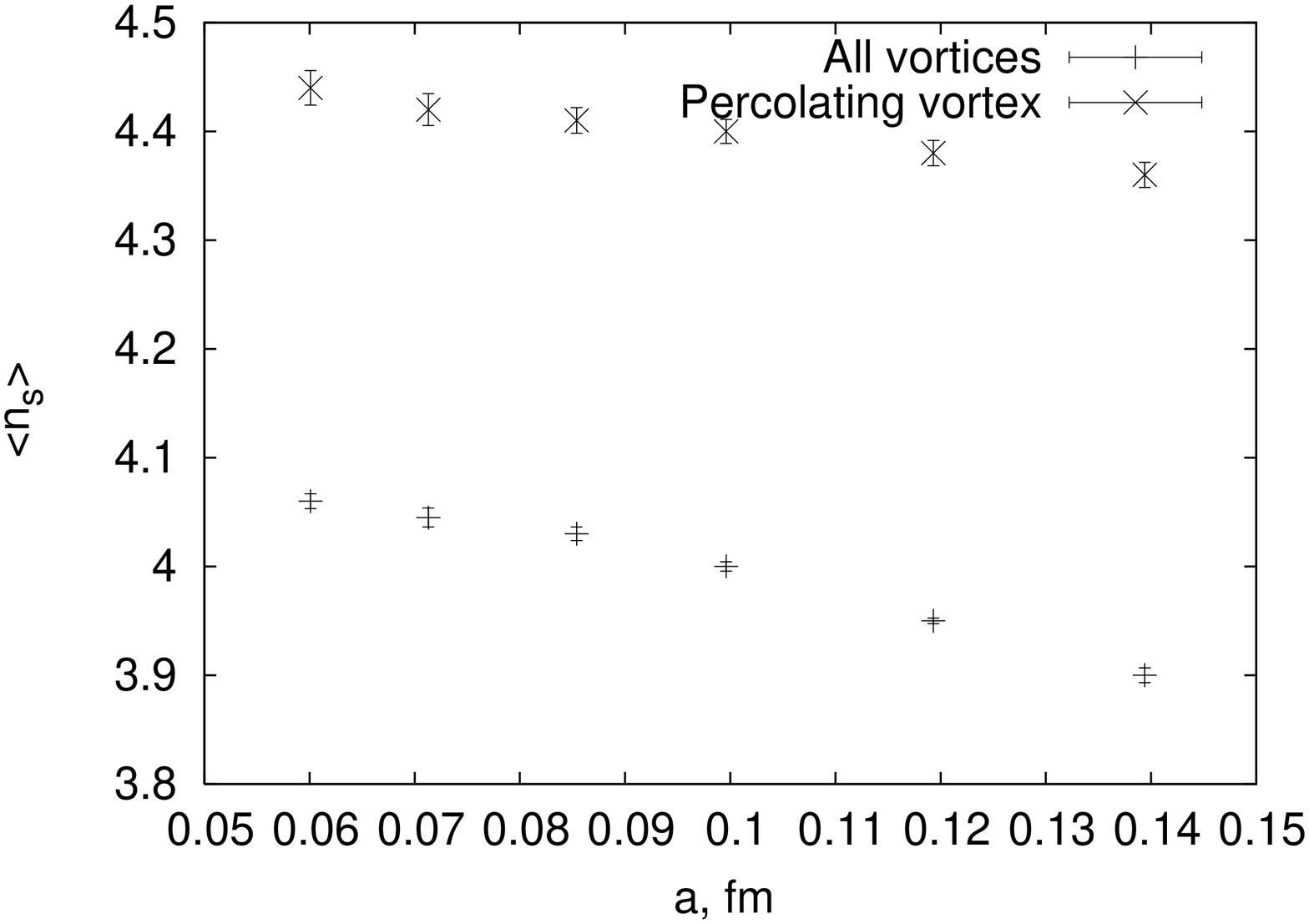}\\
  \caption{Left: average number of adjacent parallel plaquettes as the function of lattice spacing. Right: average number of neighbors per site as the function of lattice spacing.}
  \label{fig:nneigh_vs_spacing}
\end{figure}

 Quadratic divergence of surface action is a necessary condition for the physical scaling of the area of random surfaces. The reason is that for self-avoiding surfaces the number of surfaces of a given area $S$ grows as $\expa{ c \: \Lambda_{UV}^{2} S }$, where $c$ is some constant which depends on the structure of the lattice, and the sum over all surfaces can only converge if the statistical weight of each surface decreases not slower than $\expa{ -c \Lambda_{UV}^{2} S }$. In \cite{Polikarpov:03:1} it was shown that the average action associated with center vortices is indeed proportional to the total area of the vortex worldsheets and diverges quadratically:
\begin{eqnarray}
\label{Polikarpov031Fit}
\langle \: W_{eff}\lr{\Sigma} \: \rangle \approx 0.54 \: a^{-2} \: S\lrs{\Sigma} = 0.54 \: \Lambda_{UV}^{2} \: S\lrs{\Sigma}
\end{eqnarray}
Lattice simulations indicate that $\sigma_{0}\lr{a}$ is almost linear in $a$ (see Fig. \ref{fig:couplings_vs_spacing}): $\sigma_{0}\lr{a} \approx A + B \: a$, where $A = 0.05 \pm 0.02$ and $B = \lr{2.9 \pm 0.02} \: fm^{-1}$, therefore $\sigma\lr{a} = A a^{-2} + B a^{-1}$. The value of $A$ is numerically very small, and it is likely that $\sigma\lr{a}$ diverges only linearly. Thus the bare vortex tension $\sigma\lr{a}$ diverges too weakly to saturate (\ref{Polikarpov031Fit}) and to cancel the quadratically divergent surface entropy, hence quadratic divergence (\ref{Polikarpov031Fit}) in the effective vortex action should arise due to some other terms. If this action indeed depends on $K$ and $R$ only, the fit (\ref{EffVortActFit}) and the estimation (\ref{Polikarpov031Fit}) imply that curvature-dependent terms should diverge, i.e. $ \kappa\lr{a} \langle \: K \: \rangle + \gamma\lr{a} \: \langle \: \lr{R - R_{0}\lr{a}}^{2} \: \rangle \sim \Lambda_{UV}^{2}$. In order to check this we have measured the average number of parallel adjacent plaquettes $\langle \: n_{p} \: \rangle$ and the average number of neighbors per site $\langle \: n_{s} \: \rangle$ for vortex worldsheets as the functions of lattice spacing (see Fig. \ref{fig:nneigh_vs_spacing}). Extrapolation to the continuum limit shows that both $a^{2} K_{p} = 4 - n_{p}$ and $ - \: a^{2} R_{s} = n_{s} - 4$ are finite and positive as $a \rightarrow 0$, therefore $\langle \: K \: \rangle$ and $\langle \: R \: \rangle$ indeed diverge as $a^{-2}$ and $- \: a^{-2}$. As we will see  further, it is important that $\langle \: R \: \rangle$ is negative.

 Consider now the coupling constants $\kappa\lr{a}$ and $\gamma\lr{a}$. Extrapolation of the results plotted on Fig. \ref{fig:couplings_vs_spacing} to $a = 0$ shows that $\kappa\lr{a}$ is finite in the continuum limit: $\kappa\lr{0} = 0.18 \pm 0.02$, while $\gamma\lr{a}$ is proportional to $a$: $\gamma\lr{a} = C \: a = C \: \Lambda_{UV}^{-1}$, where $C = \lr{ 0.0022 \pm 0.0001 } fm$. Standard dimensional analysis shows that $\gamma\lr{a}$ should be quadratic in $a$, thus both $\sigma\lr{a}$ and $\gamma\lr{a}$ seem to have nonstandard scaling dimensions: $\: 1$ and $-1$ instead of $2$ and $-2$.

 Extrapolation of $\langle \: a^{2} K_{p} \: \rangle = 4 - \langle \: n_{p} \: \rangle$ shows that $\langle \: K \: \rangle \approx \lr{3.45 \pm 0.01} \: a^{-2}$ as $a \rightarrow 0$. According to (\ref{EffVortActFit}), the contribution of $\langle \: K \: \rangle$ to the average action density is $\langle \: \delta w_{eff}\lr{K,R; a} \: \rangle = \kappa\lr{a} \: \langle \: K \: \rangle \approx \lr{0.18 \pm 0.02} \lr{3.45 \pm 0.01} \: a^{-2} \approx \lr{0.6 \pm 0.1} \: a^{-2}$, which agrees well with the estimation (\ref{Polikarpov031Fit}). It appears that the dominating contribution to the effective vortex action $W_{eff}\lrs{\Sigma}$ comes from the term $\kappa\lr{a} K$, therefore center vortices can be effectively described as rigid tensionless strings.

 As $\gamma\lr{a}$ has scaling dimension $-1$, quadratically divergent internal curvature yields a cubic divergence in $w_{eff}\lr{K,R; a}$ unless $\langle \: R \: \rangle = R_{0}\lr{a}$. Extrapolation of $\langle \: a^{2} R_{s} \: \rangle$ to the continuum limit shows that $\langle \: R \: \rangle$ is indeed close to $R_{0}\lr{a}$ if the averaging is performed only over percolating vortex (see Fig. \ref{fig:nneigh_vs_spacing}). However, when the averaging is performed over all vortices, $\langle \: R \: \rangle - R_{0}\lr{a}$ again diverges quadratically, thus formally giving rise to a cubic divergence in $W_{eff}\lrs{\Sigma}$. This fact implies that either our measurements are flawed by some lattice artifacts (such as incomplete separation of percolating and satellite vortices) or that even higher powers of $R$ survive in the continuum limit and, consequently, our effective vortex action is not renormalizable.

\section{Percolation of center vortices}
\label{sec:percolation}

 By virtue of the Gauss-Bonnet identity $\langle \: \int \limits_{\Sigma} d^{2} \xi \sqrt{g} R \: \rangle \sim \langle \: S\lrs{\Sigma} \: \rangle \: \langle \: R \: \rangle \sim 2 - 2 \langle \: g\lrs{\Sigma} \: \rangle$, where $g\lrs{\Sigma}$ is the genus of the surface $\Sigma$. Therefore if $\langle \: R \: \rangle$ is large and negative,
the genus of vortex worldsheets is also large, i.e. there are a lot of handles on it. Since center vortices have finite density in the continuum limit, $\langle \: S\lrs{ \Sigma } \: \rangle$ and the genus $\langle \: g\lrs{\Sigma} \: \rangle$ are proportional to the total volume of physical space. This agrees well with the results of lattice simulations, which indicate that the genus of the percolating vortex is indeed very large and grows linearly with the total volume of the lattice. Thus the action of the form (\ref{EffVortActFit}) provides also a rough qualitative explanation of vortex percolation: due to the term $\gamma\lr{a} \lr{R - R_{0}\lr{a}}^{2}$, where $R_{0} < 0$, it is more advantageous to have a single connected surface with a large number of handles, so that $\langle \: R \: \rangle \sim R_{0} < 0$, than a large number of disconnected surfaces with $\langle \: R \: \rangle > 0$. It is interesting to note that the effective vortex action of the form $w_{eff}\lr{K,R; a} = \sigma\lr{a} + \kappa\lr{a} K + \gamma'\lr{a} R$ with $\gamma'\lr{a} > 0$ \cite{Buividovich:07:3} also favors negative $R$, which, however, cannot take arbitrarily large absolute values because of the positively defined term $\kappa\lr{a} K$. The decreasing of $\langle \: R \: \rangle$ should stop at some negative value $\sim R_{0}\lr{a}$, which is UV divergent, because both $\kappa\lr{a}$ and $\gamma'\lr{a}$ are dimensionless and therefore neither $K$ nor $R$ know about the $\Lambda_{QCD}$ scale. Presumably the term $\gamma\lr{a} \: \lr{R - R_{0}\lr{a}}^{2}$ in (\ref{EffVortActFit}) is just an effective description of this phenomena.

 Since $\langle \: K \: \rangle$ and $\langle \: R \: \rangle$ diverge as $\Lambda_{UV}^{2} = a^{-2}$, a typical size of geometric structures associated with center vortices should be of order of several lattice spacings. Indeed, a typical vortex configuration on the lattice, besides a single percolating vortex, contains also a large number of small satellite vortices, which consist of only several lattice plaquettes. However, such small vortices can give only positive contribution to $\langle \: R \: \rangle$, and negative UV divergent values of $\langle \: R \: \rangle$ can only be associated with percolating vortex. This implies that a typical length scale associated with percolating vortex should also be comparable with lattice spacing, which suggests that percolating vortex emerges as a result of purely statistical coalescence of small satellite vortices, just as in the case of usual percolation. On the other hand, it is difficult to explain the physical scaling of total vortex area by such purely statistical mechanism which works at distances of order of lattice spacing. It seems that tensionless rigid random surfaces and percolation theory are two complementary languages which describe one and the same phenomenon.

\section{Conclusions}
\label{sec:Discussion}

 The analysis performed above shows that the dynamics of center vortices is more complicated than that of rigid random surfaces. In particular, the bare vortex tension diverges too weakly to compensate the divergence of surface entropy, and the fluctuations of surface geometry can only be suppressed by the terms which depend on the curvature of vortex worldsheets. As a result, center vortices form some irregular geometric structure with characteristic size of order of several lattice spacings, so that $\langle \: K \: \rangle \sim \Lambda_{UV}^{2}$ and $\langle \: R \: \rangle \sim - \: \Lambda_{UV}^{2}$. This structure, however, is completely different from branched polymers which are a common problem for random surfaces with Nambu-Goto action \cite{Ambjorn:94:1}. In fact negative $\langle \: R \: \rangle$ indicates that the genus of vortex worldsheets grows with their area, which is a necessary condition for percolation. In contrast, branched polymers are usually assumed to have fixed topology \cite{Ambjorn:94:1}. It is interesting to note that two commonly assumed conditions of surface percolation, namely, vanishing of vortex tension and the existence of "false" vacuum \cite{Chernodub:07:1} are fulfilled in our case, although in a rather odd way. Percolating vortex in some sense corresponds to the "true" vacuum with nonvanishing negative $\langle \: R \: \rangle$, while small satellite vortices can be considered as perturbative excitations over this vacuum.

 This work was partially supported by grants RFBR-DFG 06-02-04010 and DFG-RFBR 436 RUS 113/739/2, RFBR 05-02-16306, and RFBR 04-02-16079, and by the EU Integrated Infrastructure Initiative Hadron Physics (I3HP) under contract RII3-CT-2004-506078.

%\bibliography{MyBibliography}

\begin{thebibliography}{1}
\bibitem{tHooft:78}
G.~{t' Hooft}, {\it On the phase transition towards permanent quark confinement},  {\em Nuclear Physics B} {\bf 138} (1978) 1 -- 25.
\bibitem{DelDebbio:98}
L.~{Del Debbio}, M.~Faber, J.~Giedt, J.~Greensite, and S.~Olejnik, {\it Detection of center vortices in the lattice {Y}ang-{M}ills vacuum},  {\em Physical Review D} {\bf 58} (1998) 094501.
\bibitem{Polikarpov:03:1}
F.~V. Gubarev, A.~V. Kovalenko, M.~I. Polikarpov, S.~N. Syritsyn, and V.~I. Zakharov, {\it Fine tuned vortices in lattice {SU(2)} gluodynamics},  {\em Physics Letters B} {\bf 574} (2003) 136 -- 140.
\bibitem{Polikarpov:03:2}
V.~G. Bornyakov, P.~Y. Boyko, M.~I. Polikarpov, and V.~I. Zakharov, {\it Monopole clusters at short and large distances},  {\em Nuclear Physics B} {\bf 672} (2003) 222 -- 238.
\bibitem{Buividovich:07:3}
P.~V. Buividovich and M.~I. Polikarpov, ``Center vortices as rigid strings.''  ArXiv:0705.3745 [hep-lat], 2007.
\bibitem{Ambjorn:94:1}
J.~Ambj{\o}rn, ``Quantization of geometry.'' Lectures presented at the 1994 Les Houches Summer School, 1994.
\bibitem{Engelhardt:00:2}
M.~Engelhardt and H.~Reinhardt, {\it Center projection vortices in continuum {Y}ang-{M}ills theory},  {\em Nuclear Physics B} {\bf 567} (2000) 249.
\bibitem{Chernodub:07:1}
M.~N. Chernodub and V.~I. Zakharov, {\it Magnetic component of {Y}ang-{M}ills plasma},  {\em Physical Review Letters} {\bf 98} (2007) 082002.
\end{thebibliography}
%\bibliographystyle{JHEP}

\providecommand{\href}[2]{#2}\begingroup\raggedright
\endgroup

\end{document}